\documentclass[final]{agujournal2019}
\usepackage{url} 
\usepackage{lineno}
\usepackage{soul}
\usepackage{gensymb}
\usepackage{amsmath,amsfonts,amsthm}
\usepackage{graphics, graphicx, caption, subcaption}
\usepackage{amsmath}
\usepackage{threeparttable}

\journalname{Space Weather}

\begin{document}

%
%

\twocolumn[
\begin{@twocolumnfalse}
\title{Relevance of Murchison Widefield Array Interplanetary Scintillation Observations to Heliospheric Transient Catalogues}

%

\authors{A. Waszewski\affil{1,2}, J. S. Morgan\affil{1}, M. C. M. Cheung\affil{3}, R. Ekers\affil{1,3}, E. Samara\affil{4,5}, S. Majumdar\affil{6}, R. Chhetri\affil{1}, N. D. R. Bhat\affil{2}, M. Johnston-Hollitt\affil{7}}

\affiliation{1}{CSIRO Space and Astronomy, P.O. Box 1130, Bentley, WA 6102, Australia}
\affiliation{2}{International Centre for Radio Astronomy Research, Curtin University, Bentley, WA 6102, Australia}
\affiliation{3}{CSIRO Space and Astronomy, P.O. Box 76, Epping, NSW 1710, Australia}
\affiliation{4}{NASA Goddard Space Flight Center, 8800 Greenbelt Rd., Greenbelt, MD, 20770, USA}
\affiliation{5}{The Catholic University of America, Washington, D.C., 20064, USA}
\affiliation{6}{Austrian Space Weather Office, GeoSphere Austria, Graz, Austria}
\affiliation{7}{Curtin Institute for Data Science, Curtin University, Bentley, WA 6102, Australia}


\correspondingauthor{Angelica Waszewski}{angelica.waszewski@icrar.org}

\begin{keypoints}
\item 57\% of catalogued interplanetary events over a 7-month period were detected in MWA IPS observations.
\item Over 55\% of IPS enhancements captured over a 7 month period were positioned out of the ecliptic plane.
\item MWA IPS observations have the ability to detect CMEs and SIRs travelling out of the ecliptic.
\newline
\end{keypoints}

%
%

%
%

\begin{abstract}
We have conducted a comprehensive comparison of interplanetary scintillation (IPS) observations taken by the Murchison Widefield Array (MWA) with several heliospheric transient event catalogues, over a time period of 7 months during solar minimum. From this analysis we have found that of the 84\% of catalogued events that have MWA IPS data available, 68\% of them appear in MWA observations. Of the enhancements first identified in IPS observations, only 58\% have a potential match with a catalogued event.
The majority of enhancements that were identified in the IPS observations were situated greater than 10$^\circ$ from the ecliptic plane. Two such features were selected for detailed analysis, connecting their solar origins to their propagation through the heliosphere.
The first of these features was created by a coronal mass ejection (CME), captured over two successive MWA observations and recorded in several catalogues. The second feature has the potential of being a stream interaction region (SIR) travelling out of the ecliptic plane. This particular SIR was not recorded in any catalogue. Thus the MWA shows promise in detecting heliospheric transients that other commonly-used techniques may overlook.
These results show the strength of the MWA in having unbridled access to the heliosphere, able to make remote observations of events far out of the ecliptic as it is not restrained to the orbits of spacecraft. We demonstrate how the inclusion of MWA IPS data can potentially boost the number of CME and SIR events that are characterised. 
\end{abstract}

\section*{Plain Language Summary}
In this paper we compared our data taken by the Murchison Widefield Array (MWA) to a series of solar event catalogues. 
Radio galaxies will twinkle due to turbulence in the solar wind, called interplanetary scintillation (IPS). By measuring this scintillation, we can get information on how the solar wind is changing. In particular, we can identify different types of heliospheric events in the solar wind, such as coronal mass ejections (CMEs) and stream interaction regions (SIRs).
We cross-matched events found in our IPS data to catalogued CME and SIR events, and found that 58\% of IPS events have a catalogued match. Doing the reverse, 57\% of catalogued events appeared in our IPS observations. 
We also conducted a detailed analysis on two specific enhancements. The first was a CME that was associated with a catalogued event, whilst the second is potentially a SIR travelling out of the ecliptic. This particular SIR was not catalogued. This shows that the MWA can detect heliospheric events that other methods may not. Therefore we show here how including MWA IPS data could potentially boost the number of CME and SIR events that we detect, especially those travelling out of the ecliptic plane. \newline

\end{@twocolumnfalse}
]

\section{Introduction}
Interplanetary scintillation (IPS) is the amplitude and phase scintillation at radio frequencies from compact sources caused by density irregularities in the solar wind \cite{Coles1978}. IPS has an extensive history of capturing and studying heliospheric structures \cite{Hewish1989}, such as stream interaction regions \cite<SIR,>{Dennison1968, Hewish1986}. Much of this seminal work that was done by the IPS community was conducted at a time when it was still not universally appreciated how solar events could drive space weather activity measured at Earth \cite{Gosling1993}, and it was partly through the use of IPS that such a connection was discovered \cite{Hewish1988}, and allowed for forecasts of geomagnetic activity \cite{Hewish1987}. Today, there are several radio telescopes and arrays around the globe that have the capabilities of detecting and measuring IPS \cite<e.g>{Tokumaru2011, Fallows2012, Naidu2015, MejiaAmbriz2010} which have been used for monitoring of the solar wind environment, and it is shown to be a useful tool in detecting and identifying individual heliospheric events over the past few decades \cite<e.g.>{Tokumaru2006, Tokumaru2023EW}, including coronal mass ejections \cite<CME; e.g.>{Jackson2007, Manoharan2010, Iwai2019, Morgan2023} and SIRs \cite<e.g.>{Bisi2010}. The characterisation of these events in IPS data, specifically the extraction of solar wind density and velocity information from the IPS signature, has been proven to improve the accuracy of large-scale solar wind models \cite{Jackson2008} and CME time-of-arrival estimations \cite{Iwai2021}.

We continue to show the advanced capabilities of the use of the phenomena of IPS with Murchison Widefield Array \cite<MWA,>{Tingay, Wayth2018}. The MWA is a low-frequency radio telescope sited at Inyarrimanha Ilgari Bundara, the Murchison Radio-Astronomy Observatory in Western Australia. It has been making continuous IPS observations for the last 6 years in the elongation range of 0.25 to 0.7\,AU, allowing us to probe large sections of the inner heliosphere using only a small number of daily observations. The cumulation of these observations have led to the first data release of the MWA IPS survey \cite{ipssurvey}, the largest catalogue of IPS sources to date, with subsequent data releases in-prep, large-scale studies of the heliospheric environment over a solar cycle \cite{waszewski2025}, and the dissection of the IPS signature to obtain information about the solar wind velocity. 

As has been shown by the MWA IPS survey, the MWA has an unprecedented number of sources available to it for conducting heliospheric studies. The majority of the historical work mentioned earlier, as well as other work that is currently being done in the field of IPS, is conducted with the use of meridional telescopes. Such telescopes have very limited sky coverage, and do not have the luxury of such high source density. It is this high density of sources, and the wide-field of view and snapshot imaging fidelity of the MWA that allows for the identification of heliospheric events, such as SIRs, in great detail with the use of a single MWA observation.

\citeA{Morgan2023} was able to track a CME from its detection in white-light coronagraph images far into the inner heliosphere with just two 5-minute observations. Following on from this work, \citeA{waszewski2023} identified several transient enhancement features whilst conducting a blind search of MWA data. In this prior study, two features captured on 04-Aug-2019 were highlighted in particular. \citeA{waszewski2023} was then able to conduct a comprehensive velocity analysis, showing that these enhancements captured in the IPS observations were travelling with an angular velocity similar to the solar wind velocities measured in-situ by the Solar Terrestrial Relations Observatory (STEREO-A) at the time of the observations, confirming that this enhancement was being created by some kind of a heliospheric event. Further data from STEREO-A assisted in identifying a small coronal hole at the centre of the solar disc a few days prior to the observations. Although only brief theorisation was given to the solar origins of these features, \citeA{waszewski2023} speculated that the enhancements were created by a SIR. In this work, we revisit this event, in an effort to answer some of the questions that stem from our earlier investigation, specifically in confirming the SIR origins.

In this study, we conducted a comparison between detected heliospheric CME and SIR events captured in MWA observations with known, catalogued events and explore the extent of overlap. From this catalogue comparison, we further the analysis of two example events which were first identified in the MWA IPS data. Both of these events were situated well out of the ecliptic plane, and represent events that were only marginally detected with commonly-used identification methods, including white-light coronagraph images and in-situ spacecraft measurements. This study presents the strengths of the MWA, and other radio telescopes, in providing full access to the inner heliosphere, with the capabilities to detect and track heliospheric events out of the ecliptic plane, where other techniques may struggle or have limited applicability.

This paper is organised as follows: Section~\ref{sect:method} provides an overview of the MWA observations and IPS nomenclature. Section~\ref{sect:compare} presents the comparison of heliospheric events detected within MWA IPS observations with catalogued events. Section~\ref{sect:indepth} presents the in-depth analysis of the two specific, example events that were identified in the MWA IPS observations, with attempts to identify their solar origins and crossover with the catalogues. A final discussion is provided in Section~\ref{sect:discuss}, where we comment on the accuracy and reliability of the MWA IPS data and the sampled catalogues, and how radio data is crucial to characterise out of ecliptic heliospheric events.

The following work makes use of several helio-based coordinate systems. An overview of those specifically used in this current work are provided in Table~\ref{tab:acronyms}. For a detailed explanation of each we direct the reader to \citeA{Thompson2006}, and for the application of these coordinate systems in \textsc{python} we point the reader to the \textsc{sunpy} documentation (\url{https://docs.sunpy.org/en/stable/reference/coordinates/index.html}).
\begin{table*}
    \centering
    \begin{tabular}{cc}
    \hline
      Coordinate System Name  & Description \\ \hline
      \makecell[l]{Helioprojective Cartesian\\coordinates (HPC)} & \makecell[l]{Observer-based coordinate frame. The origin\\is the location of the observer (the Earth). $\theta_x$ is the\\angle relative to the plane containing the Sun-Earth\\line and the Sun's rotation axis. $\theta_y$ is the angle relative\\to the Sun's equatorial plane.} \\ \hline
      \makecell[l]{Heliocentric Cartesian\\coordinates (HCC)} & \makecell[l]{A coordinate system in the heliocentric system which is\\observer-based. The origin is the centre of the Sun. The\\Z-axis is aligned with the Sun-observer line. The Y-axis\\is aligned with the component of the vector to the Sun's\\north pole that is perpendicular to the Z-axis.} \\ \hline
      \makecell[l]{Heliographic Carrington\\coordinates (HGC)} & \makecell[l]{Represents the Sun's surface in a flat, 2D format, by\\applying a Mercator projection with each Carrington\\rotation centred at a Carrington longitude of $180^\circ$. This\\longitude and latitude rotates with the Sun.} \\
      \hline
    \end{tabular}
    \caption{Description of the helio-based coordinate systems used as part of this work.}
    \label{tab:acronyms}
\end{table*}

\section{Data and Methods}
\label{sect:method}
\subsection{Observations}
 
\begin{figure}
    \centering
    \includegraphics[width=0.36\textwidth]{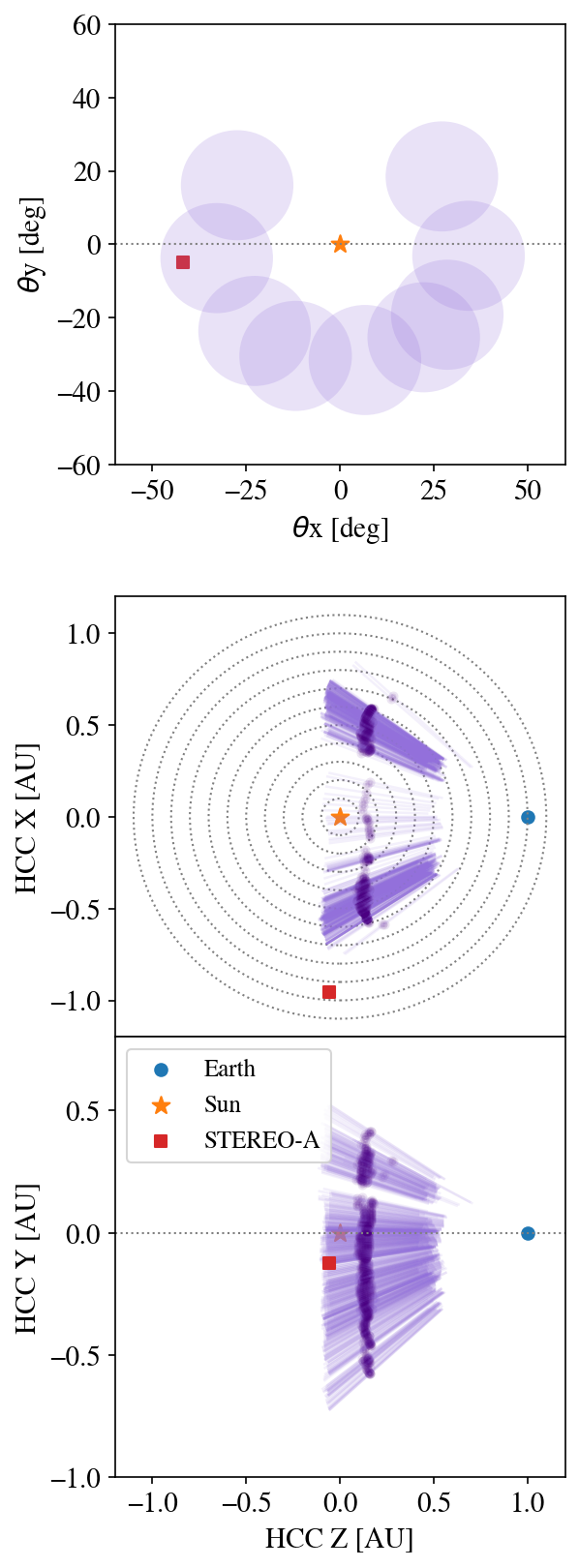}
    \caption{The sky coverage of the MWA IPS observations over a 7-month period in 2019. Top panel: Representative target fields of individual observation each in HPC, representing the sky coverage as seen by the MWA. Bottom two panels: The half-power region (light purple lines) of the line-of-sight with the piercepoint highlighted (dark purple circle marker) for the centre of all observations included in the 2019 time period. The line-of-sight is projected into Earth's orbital plane (middle panel) and into the meridional plane containing Earth (bottom panel) both in HCC. The location of Earth and STEREO-A (as found for the central observation; 30-May-2019 08:00 UTC) are added for reference. A total of 735 observations are included.}
    \label{fig:cover}
\end{figure}

This study consists of observations taken during 2019 (solar minimum), covering the period from 04-Feb-2019 to 20-Aug-2019. A subset of the observations from this observing semester have been used in several past studies, such as \citeA{ipssurvey} and \citeA{waszewski2023}. Therefore, further processing has been carried out on the remainder of the 2019 observing semester data set (Feb - Aug 2019), increasing the number of available observations to over 700. The inclusion of these additionally processed observations allows for an expanded analysis to make use of all of the heliosphere that is sampled by the MWA. The sky coverage over the sampled time range is provided in Figure~\ref{fig:cover}. We provide the sky coverage as directly seen by the MWA with representative target fields given in the top panel in HPC, while representing the heliospheric coverage as the half-power region (highlighting the point of closest approach with the Sun, further details provided in Section~\ref{sect:glevel}) of the centre of each observation in HCC taken during the sampled 2019 time period, totalling to 735 observations.

During each MWA observing semester, a series of daily observations were taken of several target fields (pointings) around the Sun. In order to target both the eastern and western limbs concentrating observations around the solar ecliptic, observations were taken at a solar elongation, $\epsilon$, of $30^\circ$, covering the solar south pole, with representative target fields given in the top panel of Figure~\ref{fig:cover} (also see Section 2.1 of \citeA{ipssurvey} and \citeA{waszewski2023}). These 10-minute observations were taken at a radio frequency of 162\,MHz and a short integration time of 0.5\,s. At this frequency the MWA's field of view (FOV) is $\sim$580 square degrees, which provides coverage from 0.25 to 0.71\,AU (roughly 54 to 153 solar radii) in the Helioprojective plane.



\subsection{g-level}
\label{sect:glevel}
Density and turbulence irregularities will cause radio sources to scintillate \cite{Coles1978}, and by quantifying this scintillation we are able to characterise these density variations within the solar wind. The measure of the level of scintillation exhibited by a source is the g-level, or scintillation enhancement factor, and is given as
\begin{equation}
    g = \frac{m_{\text{obs}}}{\text{NSI} \cdot m_{\text{pt}}}
\end{equation}
where $m_{\text{obs}}$ is the observed scintillation index of the source, and $m_{\text{pt}}$ is the expected scintillation of a point source as calculated using Equation 6 (and associated equations) detailed in \citeA{Morgan2019}, setting all parameters as for solar minimum. This same technique was used in previous heliospheric IPS studies and is described in \citeA<Section 2.2 of >{waszewski2023}, assuming the shape of the solar wind is elliptical rather than spherical, which has been shown to be the case for solar minimum \cite{Manoharan1993}.
Also included in the calculation is the normalised scintillation index \cite<NSI;>{Chhetri2018}, which dictates the fraction of the radio flux that contributes to the scintillation. With the addition of the NSI metric it will remove all effects of source structure on the scintillation observed, safely being able to compare scintillation measurements between sources. The NSI reported in the MWA IPS catalogue \cite{ipssurvey} is taken as the median over multiple observations of the same source. 
A single source measurement is actually a line-of-sight integration measurement through a large portion of the solar wind. While providing information along the whole line-of-sight, the g-level measurement has its highest sensitivity at the point of closest approach to the Sun, i.e., the piercepoint. The location of the piercepoint will vary with elongation whereas the half-power region of the line-of-sight will remain relatively consistent, as is seen in Figure~\ref{fig:cover}. Therefore, a deviation from a nominal g-level of 1 will represent a density change along the whole line-of-sight with increased sensitivity to the region closer to Earth, where a g-level of above 1 is associated with an increase in density. For the continuation of this paper, we will refer to any large increases in g-level as enhancements. 

IPS may not be the only source of variability at the observing frequencies used in this study. Ionospheric scintillation is quite prevalent at low-frequencies \cite{waszewski2022}, however, as it manifests on longer time-scales compared to IPS, it can be easily filtered out by applying a high-pass filter, whilst having minimal effects on the IPS signature itself. As described by \citeA{ipssurvey} Section 2.5, a filter with a bandpass of 0.1--0.4\,Hz is applied to each time series before computing the variance. Additional noise is contributed by the stochastic signal of the radio sources themselves \cite{morgan2021} however this is negligible even for the brightest sources we observe (and could in principle be easily subtracted from the scintillation index). Other spurious variability such as Radio-frequency Interference or solar radio bursts are easily identified in the image plane and are extremely rare in MWA IPS data. For a more extensive discussion we direct the reader to Sections 2 and 3 of \citeA{ipssurvey}.

Within a single MWA observation, we detect an order of 100 sources with g-level measurements. As there are several observations made during a single day of observing, stitching the observations together we are able to monitor 1000s of sources in a single day, therefore probing a large range of latitudes in a short time span. The simplest way of displaying this data is a g-map (e.g. Figure~\ref{fig:gmaps}; all of the g-level measurements taken over the full day of observing displayed in the Helioprojective coordinate (HPC) plane. 


\subsection{Heliospheric Transient Event Catalogues}
We have compiled a list of heliospheric CME and SIR events within our date range from several resources that are all online and publicly available (catalogues also listed in Table~\ref{tab:catalogues}); the CDAW SOHO LASCO CME catalogue \cite{Gopalswamy2009}, the CACTus LASCO CME catalogue \cite{Robbrecht2009}, and the STEREO IMPACT/PLASTIC Level 3 Event lists (\url{https://stereo-dev.epss.ucla.edu/l3_events}), including both the interplanetary CME (ICME) list \cite{Jian2018} and the SIR list \cite{Jian2019}. These catalogues were specifically chosen as they represent a range of heliospheric event classification methods, some focussing only on CMEs, while others also report SIRs. These catalogues also include a range of sensing techniques. Both the CDAW and CACTus CME catalogues are created based on white-light coronagraph images taken by the LASCO instrument \cite{brueckner1995} aboard SOHO \cite{soho}. These coronagraphs probe over all heliolatitudes, however they can only provide information very close to the Sun ($<30\,R_\odot$). This is in contrast to the in-situ measurements that are made by STEREO-A, which is located out at $\sim$1\,AU, but is limited in latitude range as its orbit will only deviate a couple of degrees from the ecliptic.

For this study, the low CME rate associated with solar minimum is a major advantage. This minimises the risk that an enhancement region identified in the MWA IPS observations is associated with the wrong or multiple catalogued events. This does also increase the number of events that may be classified as "poor" or "weak". Therefore, to ensure that only CME and SIR events that have a certain level of confidence were included in the comparison, we undertook a level of filtering for each catalogue.
\begin{enumerate}
    \item CDAW SOHO LASCO CME: Only events that were \textit{not} classified as 'Poor Event' or 'Very Poor Event' (therefore, were left as blank classifications), and had a defined velocity at 20\,$R_\odot$ were included. 
    \item CACTus LASCO CME: Only events that had an angular width of above 30$^\circ$ and had multiple velocity measurements were included. Every entry was checked in the original LASCO images.
    \item STEREO IMPACT/PLASTIC Events lists: All reported ICMEs and SIRs detected by STEREO-A were included.
\end{enumerate}
As the CDAW CME catalogue already has a system implemented where they comment on the data quality of the event, there is no need for further filtering as was required for the CACTus catalogue.


\section{Comparison with catalogues}
\label{sect:compare}
After processing all 735 MWA IPS observations in the 7-month long 2019 observing semester has resulted in 179 daily g-maps. These maps form the foundation of the comparisons between MWA IPS data with published solar event catalogues.

To compare between the catalogue events and the IPS data, we have to extrapolate the time and location of the recorded events into the MWA FOV.
The CDAW and CACTus catalogues both report a CME velocity at 20\,$R_\odot$, as well as the ejection position angle. Including this information into a simple drag-based model \cite{Vrnak2012} (outlined in Appendix A), we can estimate the CME time-of-arrival into the centre of the MWA FOV. This same process does not need to be repeated for the ICME and SIR events reported by STEREO-A. We are able to directly compare the IPS data to the in-situ measurements made by STEREO-A, as it is situated off the Eastern limb of the Sun sitting at the very edge of the MWA IPS observations (as shown in Figure~\ref{fig:cover}).
As noted previously, IPS is sensitive to a range of distances along the line-of-sight, so it is possible that a structure seen with IPS is not detected by STEREO-A in-situ. We see in Figure~\ref{fig:cover} how the piercepoint is much closer to Earth than STEREO-A is. However, IPS is likely to detect most events that were seen by STEREO-A.

From the extrapolated data, we check for each catalogued event if there is an equivalent enhancement at the estimated date and location in the MWA IPS observations via visual analysis of the daily g-maps, classifying each event into one of three categories;
\begin{enumerate}
    \item No Data: if there is no IPS data, or very poor data available ($<$20 g-level measurements in the observation) around the estimated time of arrival.
    \item No Match: if IPS data exists, but either has no enhancement at the estimated time, or shows an enhancement in a different location.
    \item Potential Match: if IPS data exists, and a plausible enhancement is contained in the observation at the estimated time of arrival. 
\end{enumerate}

We give the results of the catalogue comparison analysis in Table~\ref{tab:catalogues}.
\begin{table*}[]
    \centering
    \begin{threeparttable}
    \begin{tabular}{cccccc}
        \hline
        Event List & Total \# of Events & No Data & No Match & \begin{tabular}[c]{@{}c@{}}Potential\\Match\end{tabular} & \begin{tabular}[c]{@{}c@{}}Percentage\\Match$^\text{a}$\end{tabular} \\ \hline
        CDAW SOHO LASCO CME$^\text{b}$ catalogue & 28 & 7 & 6 & 15 & 71\% \\
        CACTus LASCO CME$^\text{c}$ catalogue & 20 & 2 & 8 & 10 & 56\% \\
        STEREO IMPACT/PLASTIC$^\text{d}$ Events & 31 & 4 & 7 & 20 & 74\% \\ \hline
    \end{tabular}
    \begin{tablenotes}
    \footnotesize
    \item[a] The percentage of events that have MWA IPS data available (excluding those events classified as No Data) that classify as a potential match.
    \item[b] Coordinated Data Analysis Workshop (CDAW) Solar and Heliospheric Observatory (SOHO) Large Angle and Spectrometric Coronagraph (LASCO) CME catalogue
    \item[c] Computer Aided CME Tracking (CACTus) Large Angle and Spectrometric Coronagraph (LASCO) CME catalogue
    \item[d] Solar Terrestrial Relations Observatory (STEREO) In-situ Measurements of Particles and CME Transients (IMPACT) / Plasma and Suprathermal Ion Composition (PLASTIC)
    \end{tablenotes}
    \caption{The total number of events included in each event list, with a breakdown of the number of those events that fall into one of the 3 classifications of No IPS Data, No Match in IPS data, and a Potential Match in the IPS data.}
    \label{tab:catalogues}
    \end{threeparttable}
\end{table*}
We see that for each individual catalogue at least 50\% of the events reported can be linked to an enhancement captured in the MWA IPS observations. Combining all the catalogued events, and ignoring any potential event overlap between catalogues, we find that of the 84\% of events that have MWA data available, 68\% have a plausible match captured in the IPS observations.
It is clear from Table~\ref{tab:catalogues} that both the STEREO-A events list and the CDAW CME catalogue have far stronger links with the IPS data as compared to CACTus. Both of these catalogues have over 70\% of the reported events that have MWA data available being linked with an IPS enhancement. 
Without conducting a more in-depth analysis, we are unable to guarantee that each matched enhancement directly corresponds to its associated catalogued event. 

\begin{figure}[t]
    \centering
    \includegraphics[width=0.8\linewidth]{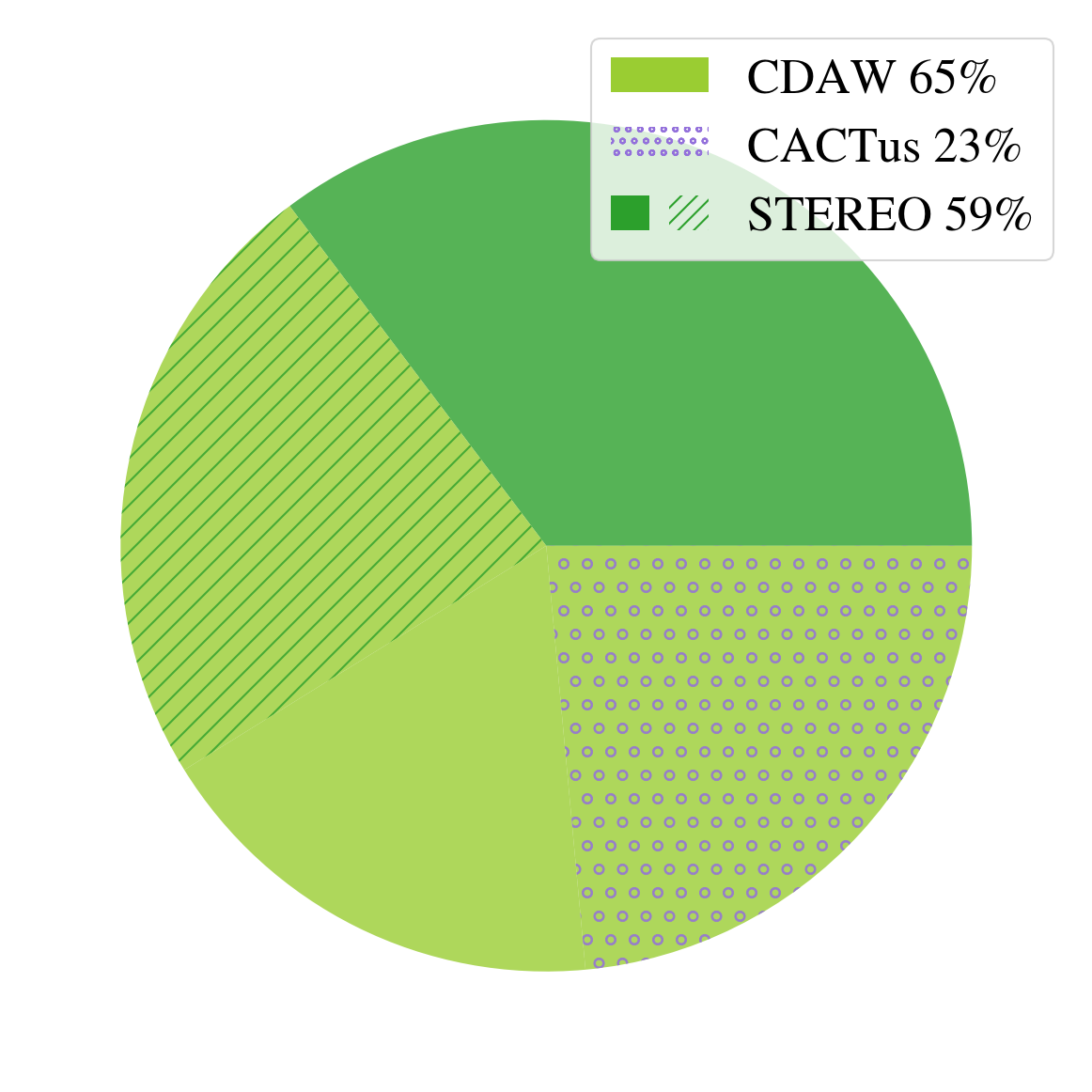}
    \caption{Summary of the 17 IPS enhancement events that had a match with a catalogued event, broken down by catalogue; CDAW SOHO LASCO CME catalogue (light green), CACTus LASCO CME catalogue (purple dots), and the STEREO IMPACT/PLASTIC Events list (dark green). The overlap between catalogues are depicted as hatching in their respective colours.}
    \label{fig:pie}
\end{figure}

In addition to first identifying an event in a series of catalogues and then extrapolating into the MWA FOV, the reverse procedure can also be conducted as a complementary investigation. This analysis involves first identifying enhancement regions in the IPS data and cross-checking whether that observation was listed during the prior catalogue comparison.
As there will be a crossover of events between the various catalogues, one IPS match can be flagged in multiple catalogues. We found that of the 32 IPS enhancement events identified via visual analysis in the $>$700 MWA observations, 17 (53\%) had been flagged as a potential match during the initial investigation. We break these 17 matches into their catalogues in Figure~\ref{fig:pie}, where it is also clear that a level of overlap exists between the catalogue themselves, especially between CACTus and CDAW. All CACTus events matched to IPS had the same event reported in the CDAW SOHO LASCO catalogue. However, no single catalogue is able to fully encompass the enhancements that are seen in the MWA IPS observations, similarly as to what was found by first identifying events in commonly-used catalogues.

Due to the nature of the dataset, we are unable to determine whether a specific IPS enhancement was created by a SIR or a CME without doing an extensive analysis on each event (such as those provided in the following Sections). However, we can break the results of the catalogue comparison with the IPS data into which type of heliospheric transient each IPS event matched too. It was found that for the 17 matches, 7 of them matched to a CME event, whilst 6 were SIRs. There were 4 IPS events which had plausible matches with both a SIR detected in the STEREO catalogue at the same time as a CME from the CDAW catalogue. Without doing an extensive analysis on each of the 17 matches, we are not able to confidently say that the associated event created the IPS enhancement, in particular to separate whether it was a SIR or a CME that created the enhancement for the 4 events that matched with both. It must also be noted that these results will be biased towards CME events as there are 3 CME catalogues compared to one for SIRs. 


\section{Identification of IPS Events}
\label{sect:indepth}
\begin{figure}
    \centering
    \begin{subfigure}[]{\linewidth}
        \includegraphics[width=\textwidth]{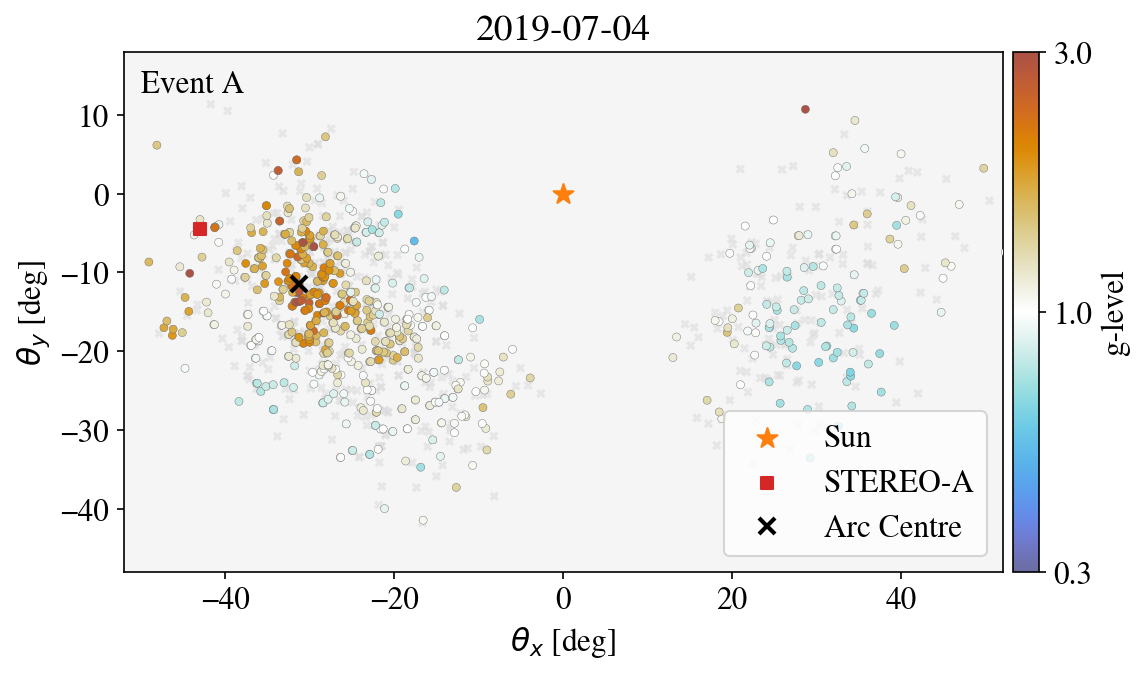}
        \includegraphics[width=\textwidth]{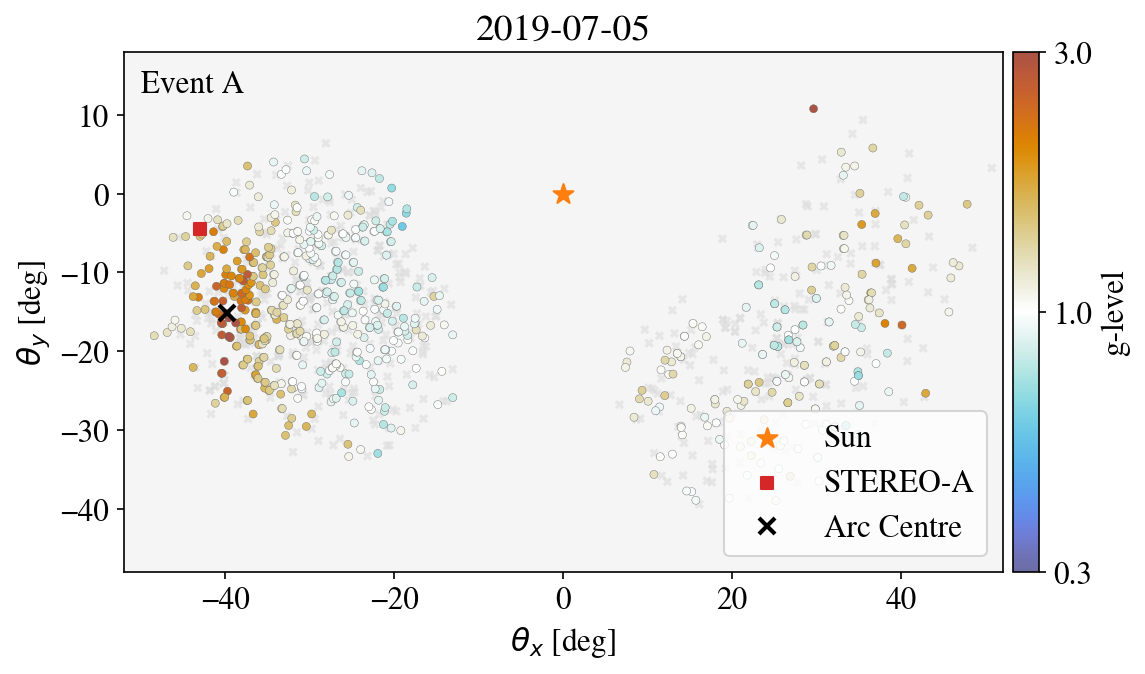}
        \caption{Event A}
    \end{subfigure}
    \begin{subfigure}[]{\linewidth}
        \includegraphics[width=\textwidth]{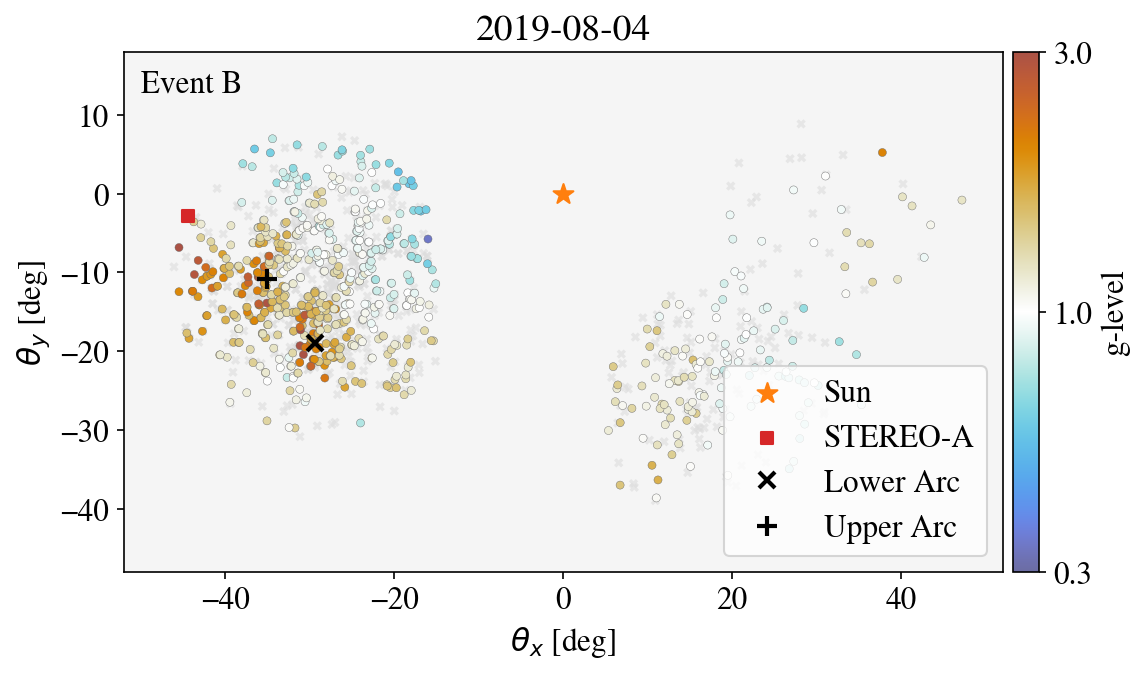}
        \caption{Event B}
    \end{subfigure}
    \caption{Daily g-maps in HPC for the two example interplanetary events chosen for further analysis; Event A (top panel) observed on 04-Jul-2019 and 05-Jul-2019, and Event B (bottom panel) observed on 04-Aug-2019. For each event, the centre of their corresponding structure/s are marked as either a cross or plus.}
    \label{fig:gmaps}
\end{figure}

It is clear from the prior analysis, that there is a relatively large discrepancy between the enhancement events identified in the MWA IPS observations and the catalogued CME and SIR events. To assess where this discrepancy is being introduced from, we selected two example MWA IPS enhancement events to conduct an in-depth analysis into their solar origins and how they crossover with the catalogues. From a preliminary investigation undertaken whilst comparing with the catalogues, it is clear that only one of these two events have a potential match within the catalogues.
The first event (henceforth referred to as Event A) was selected as it represents an agreement between the MWA IPS data and the catalogues. Shown in Figure~\ref{fig:gmaps}a is the main enhancement region of this event (the centre of which is marked with a cross) that appears to persist over two days, with the majority of the enhancement being captured in the East (E) pointing on both days in observations taken on 04-Jul-2019 07:17 UTC and 05-Jul-2019 07:17 UTC. This event appears to be correlated with a CME event that was captured in the LASCO coronagraph white-light images, and was recorded in both the CACTus and CDAW CME catalogues. We present a detailed analysis into the likelihood of Event A being a CME in Section~\ref{sect:cme}.
The second event (henceforth referred to as Event B) appears in the MWA IPS observations exactly a month after Event A, and is the same event as was identified by \citeA{waszewski2023}. Although being confirmed by \citeA{waszewski2023} that this enhancement was being created by some kind of heliospheric event, concluding that it was most likely a SIR, there was no catalogued event it could be matched with. Event B, shown in Figure~\ref{fig:gmaps}b, has two main regions of enhancement; the lower and upper arcs (centres designated as a cross and plus respectively), both of which are fully encompassed by the South East (SE) pointing observation taken on 04-Aug-2019 04:56 UTC. In Section~\ref{sect:sir}, we further investigate the nature of Event B, and conclude that both enhancement regions are being created by SIRs, one of which was not intercepted by STEREO-A or any other satellites.

\subsection{Event A -- CME}
\label{sect:cme}
While conducting the catalogue comparison, we extrapolated the time of arrival of various CMEs into the MWA FOV ($\sim0.5$\,AU). For several CMEs that were observed by SOHO LASCO, there were enhancements present in the IPS observations at their estimated time of arrival. One such CME, that had been matched to the enhancements captured in Event A, entered the LASCO C2 field of view on 01-Jul-2019 19:12 UTC, and was recorded in the CDAW SOHO LASCO CME catalogue as well as the CCMC DONKI database (\url{kauai.ccmc.gsfc.nasa.gov/DONKI/view/CME/14859/1}). Feeding the reported velocity of 384\,km/s (as stated by CCMC DONKI) into the simple drag-based model, we expected the CME to reach the centre of the MWA FOV on 04-Jul-2019 at 01:13 UTC. This estimated time of arrival only differed by 6 hours to the time of the MWA observations of Event A, taken at 07:17 UTC. Travelling at the reported velocity, it is expected that the CME would remain in the MWA FOV for over 24 hours.
Whilst using the drag-based model, we also assume that the CME is travelling in the plane-of-the-sky of the observation, more specifically the piercepoint. 
The source region of this CME was identified by combining observations from various passbands of the STEREO/EUVI-A and it was found to be located at a longitude of -86$^\circ$ and latitude of -9$^\circ$ in Stonyhurst coordinates (for details on the source region identification, please see \cite{majumdar2023}). This confirms that is was a limb event with respect to the sun-earth line, therefore it is not expected to be travelling in the plane of the observation. As the highest sensitivity region (the piercepoint and the half-power region) of the observations were much closer to the Earth, there may be slight deviations in the MWA FOV arrival time.

Following the techniques developed by \citeA{waszewski2023}, we calculated the angular velocity of the enhancement to be $313\pm10$\,km/s between the two E observations (04-Jul and 05-Jul) associated with Event A. This velocity is slower than that recorded by CDAW and DONKI entries (see Table~\ref{tab:param}). This discrepancy is most likely due to the aforementioned projection effects caused by the CME travelling not directly towards the piercepoint.

We were able to mitigate potential errors caused by projection effects by simulating the CME's evolution and then projecting into the same plane as the MWA observations. This was achieved using the implementation of the HUXt model \cite<Heliospheric Upwind Extrapolation with time dependence,>{Owens2020, Barnard2022} in \textsc{python} \cite{HUXt}. HUXt is a lightweight solar wind model that simulates the evolution of CMEs and other trace particles through the heliosphere. It is built to be computationally efficient and inexpensive, making it incredibly accessible. HUXt generates and simulates solar wind velocities in both 1D and 2D, but with a wrapper is able to run multiple latitudes in parallel, providing solar wind velocity information in 3D. It should be noted that HUXt cannot provide plasma density information. A more direct comparison with IPS data would be possible by using a full magnetohydrodynamic (MHD) simulation as to compare density and g-level measurements. Such a comparison is left for future work, where here we use HUXt, which is instead an emulated 3D MHD simulation.

We defined the base solar wind velocity profile using the inbuilt functionality of the HelioMAS model \cite{Riley2001} for a specified Carrington rotation (for this particular event, 2219), and then injected a CME with the parameters outlined in Table\,\ref{tab:param}.
\begin{table}[]
    \centering
    \begin{threeparttable}
    \begin{tabular}{lc}
        \hline
        Parameter & Value \\ \hline
        Initiation Time$^\text{a}$  & 02-Jul-2019 04:37\,UTC \\
        Longitude$^\text{b}$ & -90$^\circ$ \\
        Latitude$^\text{c}$  & 115$^\circ$ \\
        Velocity$^\text{d}$  & 493\,km/s \\
        Angular Width$^\text{e}$ & 44$^\circ$ \\
        Thickness$^\text{f}$ & 3\,$R_\odot$ \\ \hline
    \end{tabular}
    \begin{tablenotes}
    \footnotesize
        \item[a] Time at 21.5\,$R_\odot$, reported by CCMC DONKI
        \item[b] In Stonyhurst coordinates (i.e. from Earth's perspective), reported by CCMC DONKI
        \item[c] Position angle from solar North, reported by CDAW SOHO LASCO CME catalogue
        \item[d] Velocity at 20\,$R_\odot$, reported by CDAW SOHO LASCO CME catalogue
        \item[e] Reported by CDAW SOHO LASCO CME catalogue
        \item[f] Default in HUXt
    \end{tablenotes}
    \caption{The initial CME parameters injected into the HUXt simulation.}
    \label{tab:param}
    \end{threeparttable}
\end{table}
The inner boundary of a HUXt simulation is set at 30\,$R_\odot$, therefore we had to define the CME parameters as close as possible to this boundary. Due to limitations in the simulation as well as the sparsity of information available for this CME, we were only able to define the parameters of the CME up to 21.5\,$R_\odot$ (where possible using the available catalogues). 

\begin{figure*}
    \centering
    \includegraphics[width=0.9\textwidth]{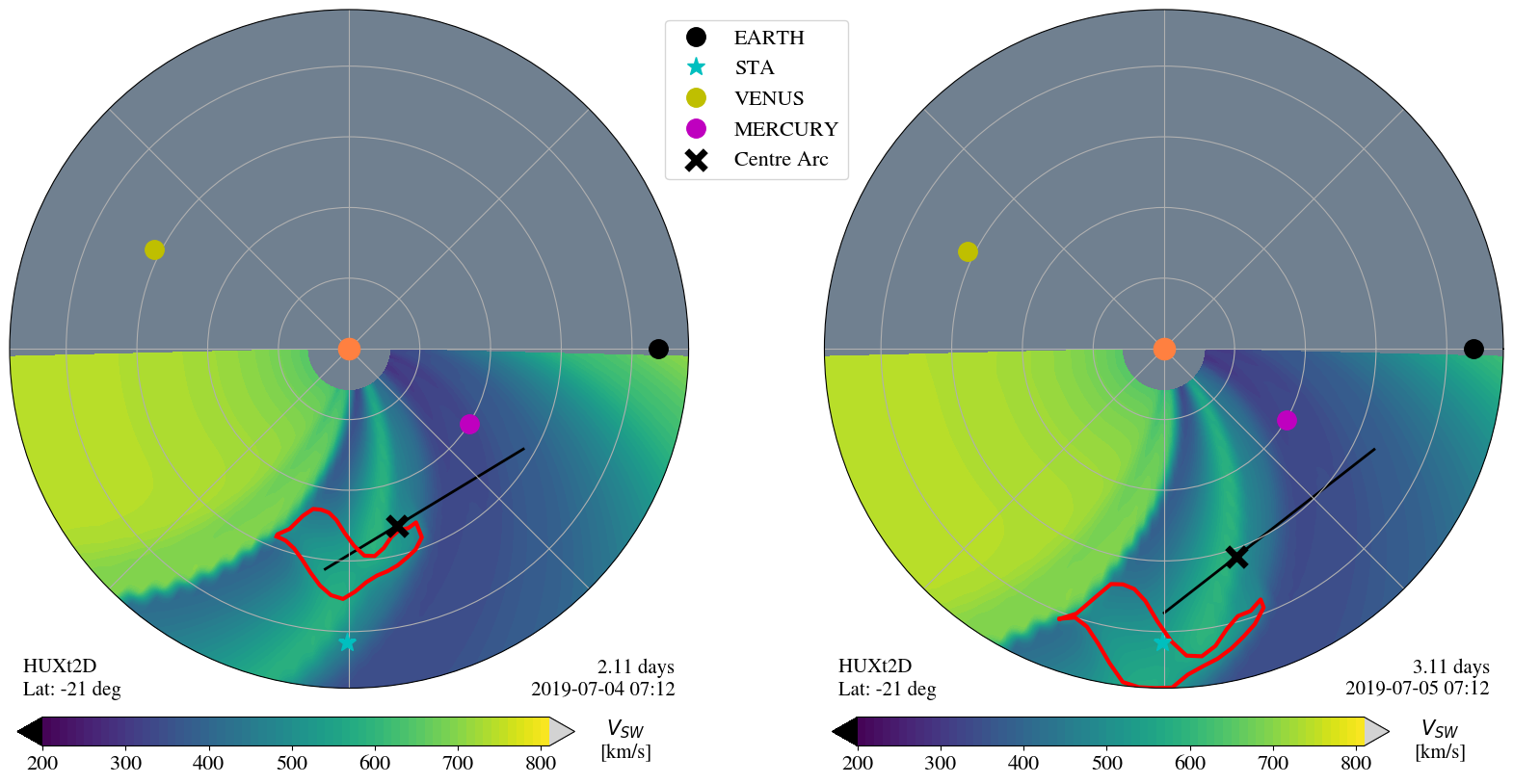}
    \includegraphics[width=0.41\textwidth]{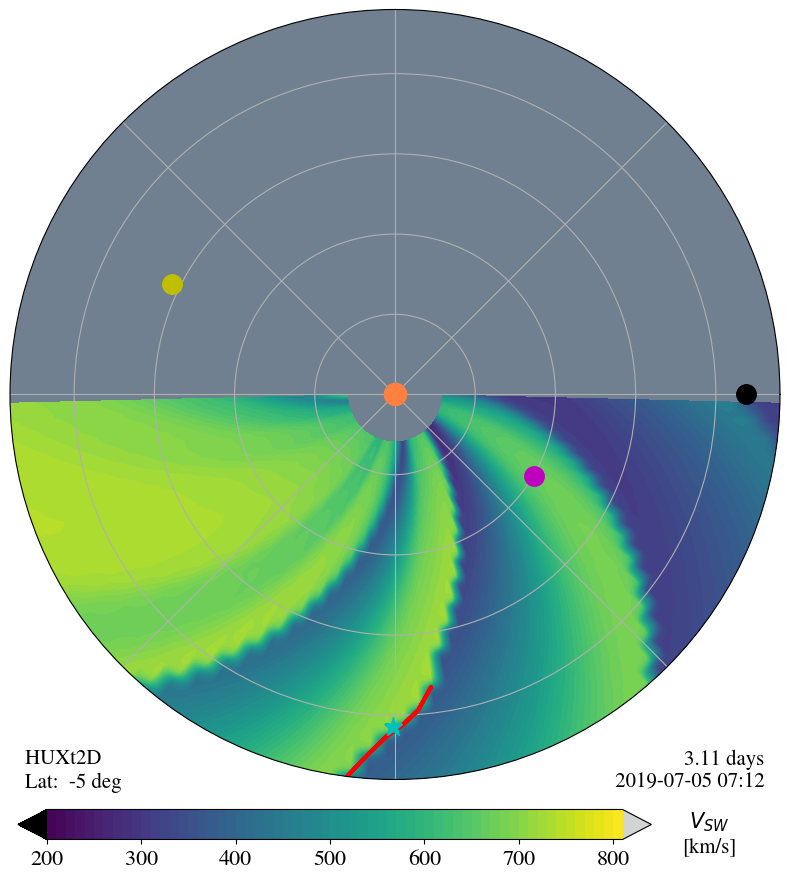}
    \caption{All panels: A single latitude, and a single timestep from a HUXt simulation. The radial axis is distance from the Sun, but rather than just an equatorial cut, the plot is a projection of the surface of a cone (with the Sun at the apex) into the 2D plane. The location of the simulation CME front is outlined in red. The half-power region of the line-of-sight of the centre arc is shown as a black line, with the piercepoint marked. The base velocity profile of the solar wind velocity (km/s) is given as a colour gradient. Top panels: Latitude of -21$^\circ$, with the left panel showing the timestamp (04-Jul-2019 07:12 UTC) nearest the time of the first observation of Event A (04-Jul-2019 07:17 UTC), whilst the right panel shows the timestamp (05-Jul-2019 07:12 UTC) of the second observation (05-Jul-2019 07:17 UTC). Bottom panel: Latitude of -5$^\circ$ to align with STEREO-A (STA in legend), at the same timestep as top right.}
    \label{fig:huxtec}
\end{figure*}

By then running a 5-day long 3D HUXt simulation, we were able to extrapolate the location of the CME at any point of its expansion by tracking test particles as they move through the simulation. HUXt has the additional functionality of creating 2D latitudinal cuts through the 3D simulation, where its output is the ambient solar wind as set by HelioMAS as the base map, with the location of the CME as found from several ensemble runs outlined on top. We show in Figure~\ref{fig:huxtec} the snapshot at a latitude of -21$^\circ$ which closely aligns with the latitude of the centre of the enhancement of Event A. Projecting the half-power region of the line-of-sight, we see the piercepoint sits inside the simulated CME for the first observation of Event A, whereas in the second appearance, the CME appears to travel further out with the edges of the line-of-sight sampling the CME. 

\begin{figure*}
    \centering
    \includegraphics[width=0.9\textwidth]{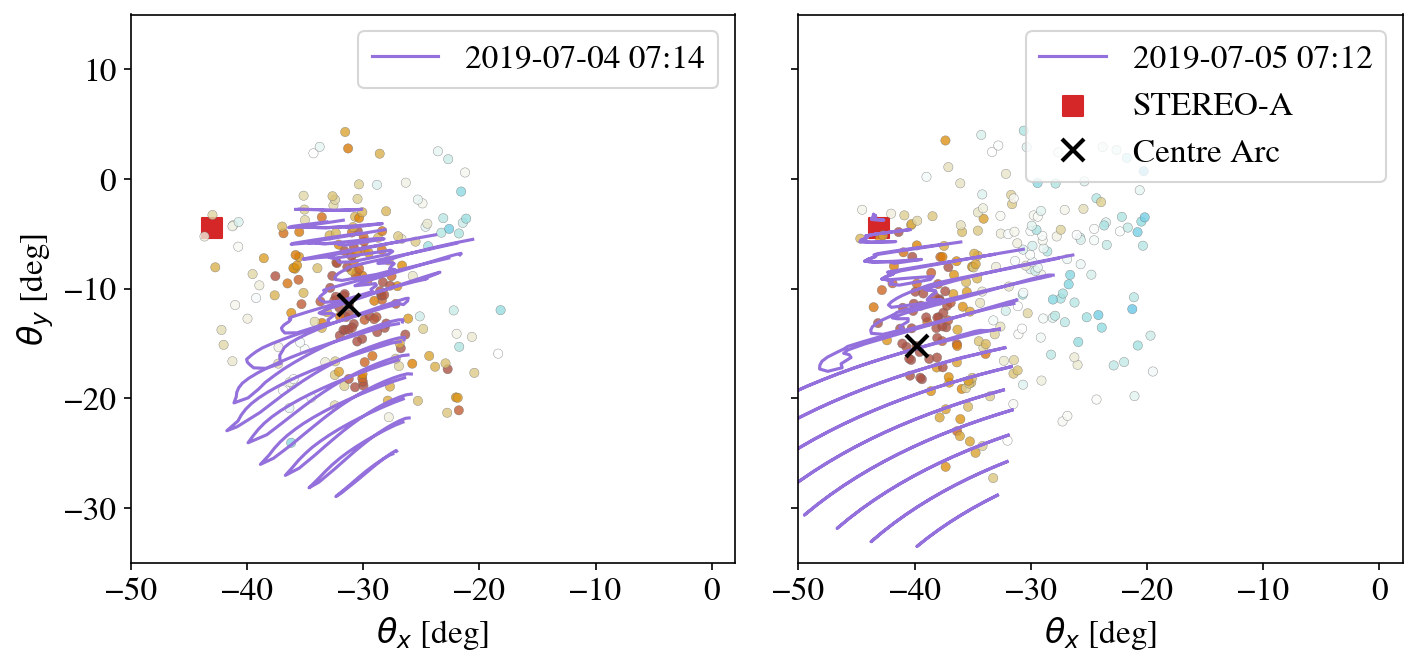}
    \caption{The location of the CME front from a 3D HUXt simulation projected into the HPC plane, outlined in purple, as compared to the g-maps of Event A. The left panel shows the HUXt simulation at the timestamp (04-Jul-2019 07:14 UTC) nearest the time of the first observation of Event A (04-Jul-2019 07:17 UTC), whilst the right panel shows the timestamp (05-Jul-2019 07:12 UTC) of the second observation (05-Jul-2019 07:17 UTC). The centre point of the enhancement and the projected location of STEREO-A are included for reference.}
    \label{fig:huxt3d}
\end{figure*}

We then directly compared the location of the g-level enhancements of Event A with the CME by projecting the outline of the HUXt model CME front into the Helioprojective coordinate plane. Once again in the projected plane, the area that the CME is contained in overlaps with the IPS enhancements at both timestamps, as shown in Figure\,\ref{fig:huxt3d}. This analysis also suggested that it is quite likely that the two enhancements seen across the two days are related. 

\begin{figure}[t]
    \centering
    \includegraphics[width=0.5\textwidth]{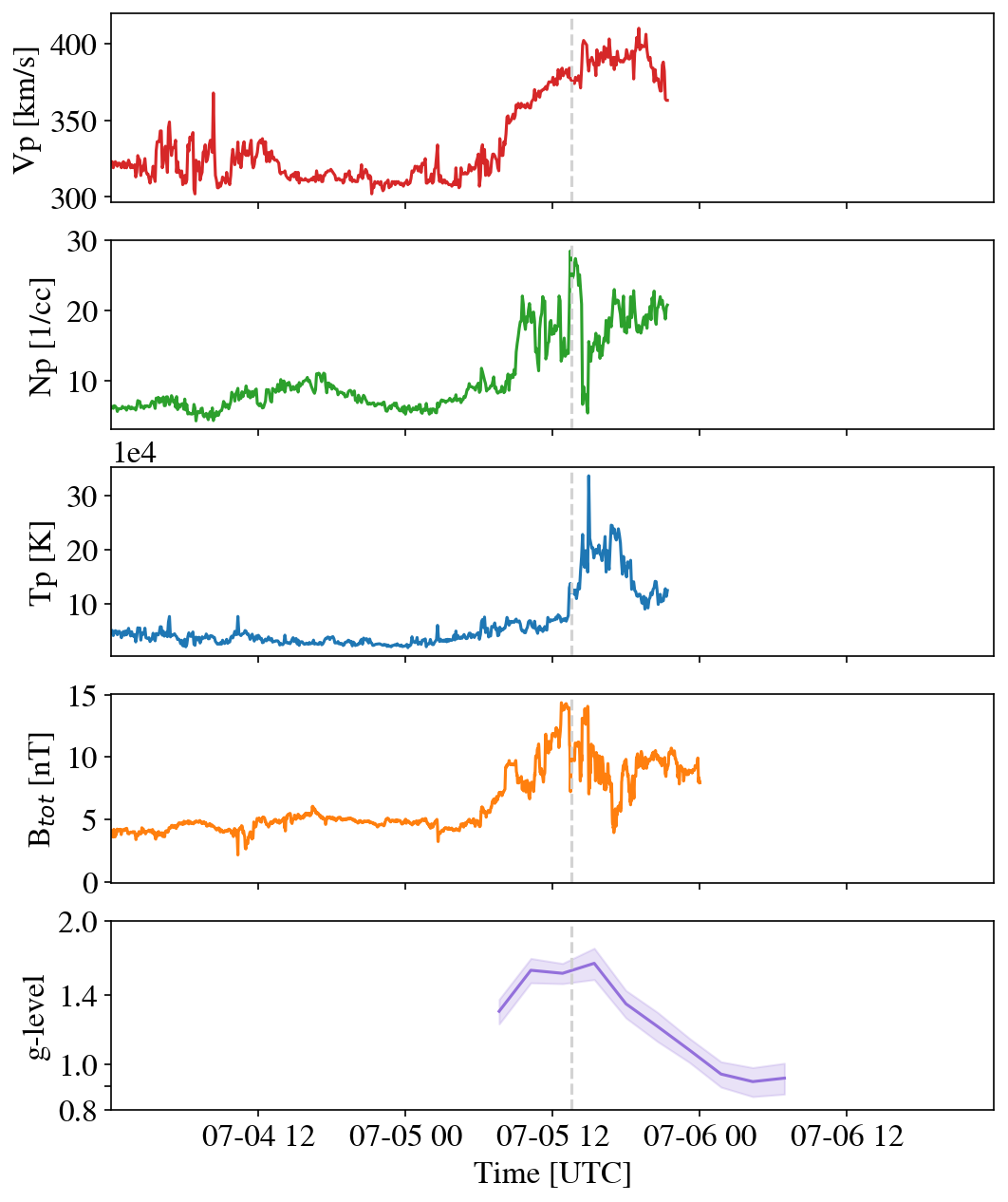}
    \caption{Level 2 magnetic field and PLASTIC solar wind plasma data from STEREO-A, including the proton velocity, Vp (red), proton number density, Np (green), proton temperature, Tp (blue), and total magnetic field, B$_{tot}$ (orange). Also included is the g-levels projected to the location of STEREO-A (purple). The dashed line is used for reference and is based on the peak of the enhancement of the second appearance of Event A.}
    \label{fig:stereo}
\end{figure}

As evident from in the lower panel of Figure~\ref{fig:huxtec}, the simulated CME appears to impact STEREO-A on 05-Jul-2019, although there is no record of this CME in the STEREO-A ICME Events list. We therefore checked the in-situ measurements made by STEREO-A's PLASTIC and magnetometer instruments to identify whether or not STEREO-A was in fact impacted. In Figure~\ref{fig:stereo} we show a portion of the Level 2 magnetic field and PLASTIC solar wind plasma data (accessed from \url{https://stereo-dev.epss.ucla.edu/l2_data}). These particular parameters shown in Figure~\ref{fig:stereo} (e.g. the solar wind velocity, density, temperature, and total magnetic field) are commonly used to identify in-situ SIR and CME events, and are also used to identify events that are reported in the STEREO-A ICME and SIR Events lists. We also include the mean g-level with its uncertainty as would be measured from the perspective of STEREO-A. Taking only g-level measurements made between -10$^\circ$ and 0$^\circ$ in $\Theta_y$ and assuming completely radial flow with a CME velocity of 384\,km/s, we propagated the g-levels out through the heliosphere and thereby estimate a g-level timeseries at the location of STEREO-A. We see that at the time of the peak g-level enhancement, STEREO-A measured an increase in proton number density (Np) and total magnetic field strength (B$_{tot}$), followed by an increase in proton temperature (Tp) and accompanied by a slow increase in proton velocity (Vp).

Despite the fact that this event is not reported in the STEREO-A Events lists, the in-situ measurements possess some of the characteristics that are attributed to a SIR; a steady and slow increase of Vp, an increase in Np (unable to see if Np decreases as no data is available after 04:00 UTC), an increase in Tp, and a compression of the magnetic field \cite{Jian2019}. To classify as a SIR event in the STEREO events list it must meet at least 5 of the criteria stated by \citeA[ and references therein]{Jian2019}, and potentially due to data loss there may be issues associated with fully characterising the proton number density fluctuations, leading to this event being a rejected candidate. We see in Figure~\ref{fig:huxtec} and Figure~\ref{fig:huxt3d} that only the very top of the CME grazed STEREO-A, but there appears to be a high-speed stream (HSS) that propelled the CME within the solar wind profile provided by HelioMAS. This HSS also appears to impact STEREO-A at the same time as the CME, as well as at the time STEREO-A in-situ measurements show a potential SIR impact. 
As the trajectory of the enhancement in the IPS observations appears as if it is travelling away from the solar equator, it may be that STEREO-A intercepted this propelling SIR, whilst the IPS observations have captured the lower latitude CME that was detected first in the white-light coronagraph images. Although the CME is relatively faint in the coronagraph images, the CME shock front may be being enhanced in the IPS by the interaction with the propelling HSS.

\subsection{Event B -- SIR}
\label{sect:sir}
Although there is strong agreement between the IPS observations and other common methods of heliospheric event identification for Event A, there is a large percentage of events identified in MWA IPS observations that have no potential match with a traditionally catalogued event. One such event is Event B. As has already been stipulated by \citeA{waszewski2023}, Event B is clearly a heliospheric event, most likely being created by a SIR. They identified the potential origins of this enhancement as a coronal hole that was visible from STEREO-A SECCHI instrument several days prior to the MWA observation of Event B. This was in addition to a HSS that was reported by STEREO-A at a similar time.
Despite both Events A and B appearing similar in nature within the IPS observations, there was no catalogued event that matched with Event B. Therefore, we conducted a similar analysis as was done on Event A, as to perform a like-for-like comparison, on why Event A was catalogued in CDAW and DONKI, whilst Event B was not.

\begin{figure}[t]
    \centering
    \includegraphics[width=0.48\textwidth]{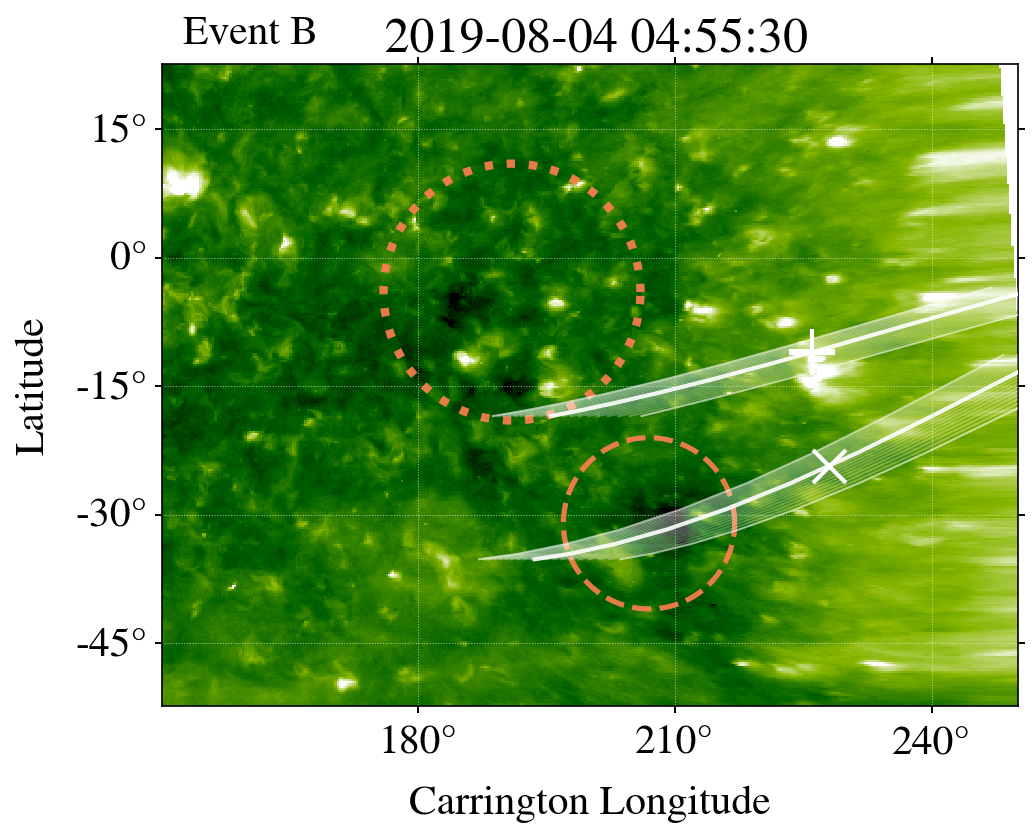}
    \caption{The half-power region of the line-of-sight (solid white line), with the piercepoint marked (cross for the lower arc, plus for the upper arc), for the lower and upper arcs in the observation of Event B. The line-of-sight is projected back onto the surface of the Sun (represented by a HGC plane) using an assumed HSS velocity of $460\pm100$km/s (uncertainty bounds shown in shaded region). A STEREO-A EUVI $195\AA$ snapshot is used as the base image, with the time of the image being as close to the time of the SE MWA IPS observation that contained both arcs (04-Aug-2019 04:56\,UTC). Also highlighted are the two coronal holes; the equatorial coronal hole (dotted orange circle) identified as creating a HSS that was identified by STEREO-A PLASTIC, and a second low-latitude coronal hole (dashed orange circle). The location of the coronal holes was defined from the EUVI $195\AA$ snapshot image at the estimated lift-off time (01-Aug-2019 23:55\,UTC).}
    \label{fig:carr}
\end{figure}

A few hours prior to the observations of Event B, at 04-Aug-2019 00:00 UTC, STEREO-A's IMPACT and PLASTIC instruments detected a HSS which was reported in the STEREO-A events list, and is also catalogued in the CCMC DONKI database (\url{kauai.ccmc.gsfc.nasa.gov/DONKI/view/HSS/14938/2}). It was found that this HSS had a peak solar wind speed of 460\,km/s, and was identified as originating from a coronal hole at the Sun's disc centre as viewed from STEREO-A EUVI \cite{Wuelser2004} instrument on 01-Aug-2019. Several days later, at the time of the MWA IPS observation, this equatorial coronal hole is still visible to STEREO-A EUVI, along with an additional low-latitude coronal hole also visible in the EUVI $195\AA$ image, both highlighted in Figure~\ref{fig:carr}. As can be seen in Figure~\ref{fig:gmaps}b, Event B contains two enhancements, both a lower and upper arc. These enhancements will be treated separately in this study.

Assuming that all outflows from both the equatorial and low-latitude coronal hole have the same velocity as was recorded by STEREO-A, we estimated the amount of time it would take for a HSS to reach the piercepoint for the lower and upper enhancements. Implementing the same basic drag-based model that was used to estimate the CME time of arrival for Event A, we traced the solar wind that could have created both the upper and lower enhancements back to the surface of the Sun to estimate a lift-off time. For this estimation we assumed an ambient solar wind speed, which was also extracted from measurements taken by STEREO-A's PLASTIC (the lowest solar wind speed recorded preceding the HSS) and was found to be 350\,km/s. 
A HSS travelling with a velocity of 460\,km/s through the ambient solar wind would take 53 hours to reach the piercepoint of the centre of both enhancements, therefore marking the estimated lift-off time as 01-Aug-2019 23:55\,UTC. By incorporating the rotation of the Sun, we projected the full line-of-sight measurement onto the solar surface using the same technique. As solar rotation is incorporated, we account for the time delay introduced for a constant velocity solar wind to travel further to reach more extreme sections of the line-of-sight (further from the piercepoint). As we observed directly off the eastern limb of the Sun, we used images taken by STEREO-A's EUVI instrument for the comparison.  As we account for the solar rotation, we use a snapshot as close to the time of the SE MWA IPS observation (04-Aug-2019 04:56\,UTC) as possible as the base EUVI image projected into the HGC frame, and we then projected the half-power region of the line-of-sight (marking the piercepoint) for both the upper and lower arc in Event B with results shown in Figure~\ref{fig:carr}. To provide an indication of the effect of any uncertainty on the assumed velocity, we also include the uncertainty region for $\pm100$\,km/s. We also highlight in Figure~\ref{fig:carr}, the location of both the equatorial and low-latitude coronal holes. The location of both was visually determined from a STEREO-A EUVI $195\AA$ image taken at the estimated lift-off time.

As shown in Figure~\ref{fig:carr}, a large section of the half-power region of the line-of-sight of the lower arc crosses directly over the low-latitude coronal hole. Although much further from either coronal hole, the upper arc's half-power region appears to be near the boundary of the equatorial coronal hole, appearing to sample the space between the two.

To further track where a SIR might be created, we repeated the HUXt 3D simulation, where now instead we injected two separate HSSs originating from each of the coronal holes, and project the location of the HSSs into the plane of the MWA observations. The location and dimension of each coronal hole is found from the STEREO-A EUVI images at instead the estimated lift-off time (01-Aug-2019 23:55\,UTC), and acted as the starting point of the HSS injected into the HUXt model. Once again to account for the inner boundary of 30\,R$_\odot$ that is incorporated in the HUXt models, we adjusted the starting longitude of each coronal hole by 7$^\circ$ to accommodate the average coronal hole rotation rate \cite{Bagashvili2017}.
As we defined our own HSS environment, we do not assume any pre-defined base solar wind velocity profile, instead opting on creating our own based on in-situ measurements made by STEREO-A to estimate the speed of high-speed and ambient solar winds. This will create a continuous and uniform ambient solar wind that the HSSs are then injected into. 

\begin{figure*}
    \centering
    \includegraphics[width=0.9\textwidth]{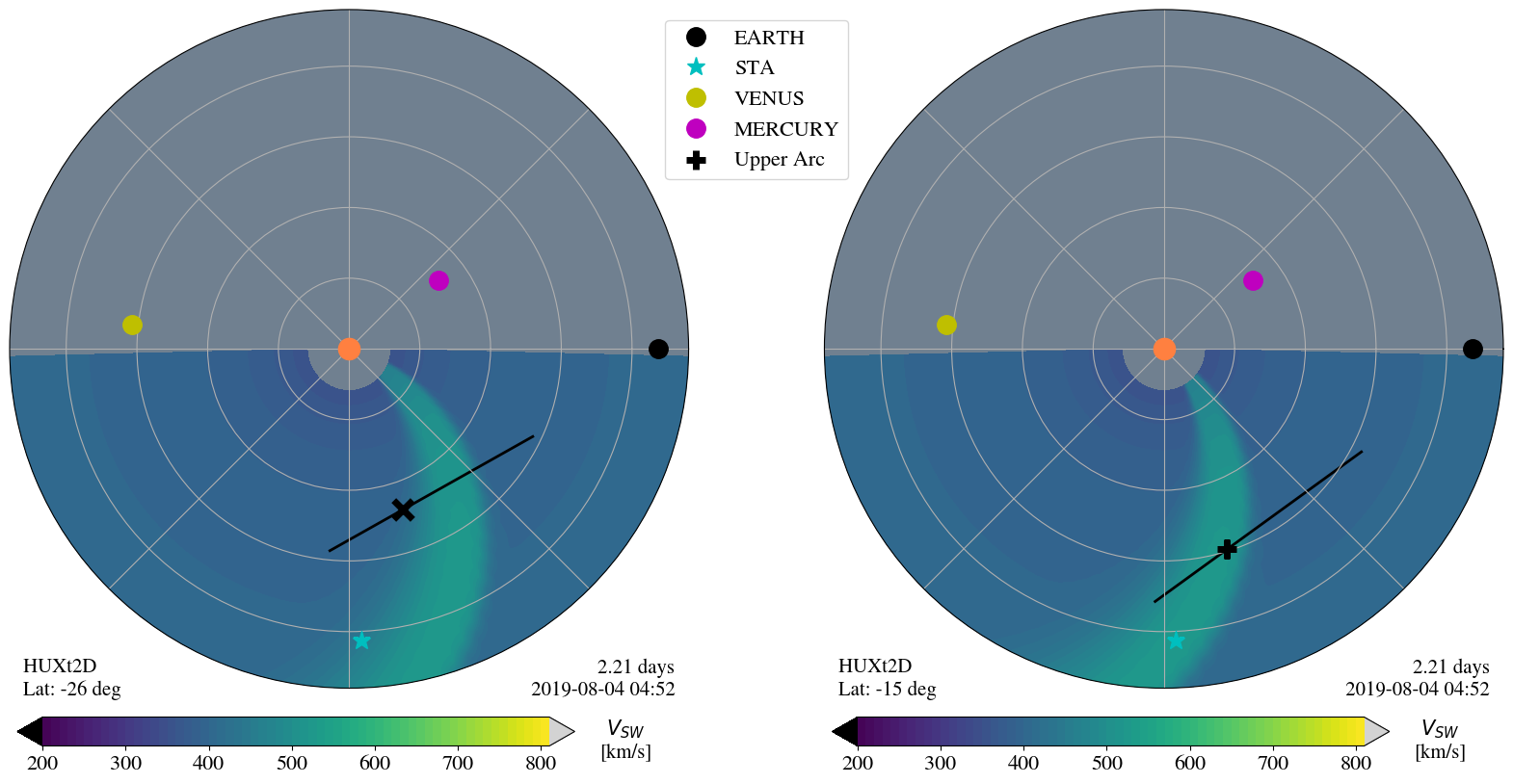}
    \includegraphics[width=0.41\textwidth]{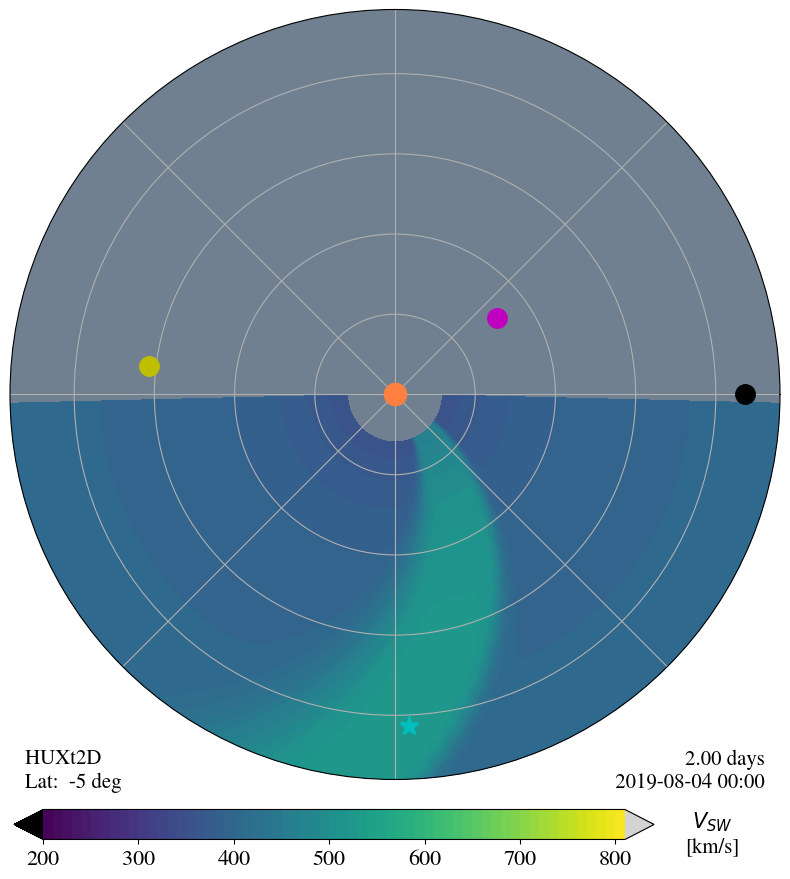}
    \caption{All panels: A 2D latitudinal cut made of the 3D HUXt simulation. The half-power region of the line-of-sight of the centre of the arcs are shown as a black line, with the piercepoint marked. The base velocity profile of the solar wind velocity (km/s) is given as a colour gradient. Top panels: Cuts made at the timestep (04-Aug-2019 04:52 UTC) nearest to the observation made of Event B (04-Aug-2019 04:56 UTC). Left panel shows a latitude of -26$^\circ$ to align with the lower arc, whilst the right panel shows a latitude of -15$^\circ$ to align with the upper arc. Bottom panel: Latitude of -5$^\circ$ to align with STEREO-A (STA in legend) at the timestep nearest to the in-situ HSS observation made by STEREO-A (04-Aug-2019 00:00 UTC).}
    \label{fig:huxtstream}
\end{figure*}

Whereas the HSS velocity was assumed to be a single value, we instead run an ensemble of HUXt simulations using a distribution of velocities. Two normal distributions were created both for the high- and ambient velocity, with a mean of the previously assumed velocities (high-speed of 460\,km/s, ambient of 350\,km/s) and a standard deviation of 50\,km/s, as found from the scatter in the solar wind velocity recorded by STEREO-A PLASTIC. The HUXt simulation was run over 50 ensembles, each sampling an HSS and ambient solar wind speed from the respective velocity distribution. To visualise where the streams are in the heliosphere we ran a base HUXt simulation with the mean values of the high-speed and ambient solar wind, with the resulting streams at the arc's respective latitudes shown in Figure~\ref{fig:huxtstream}, as well as at the latitude of STEREO-A.

\begin{figure}[t]
    \centering
    \includegraphics[width=0.5\textwidth]{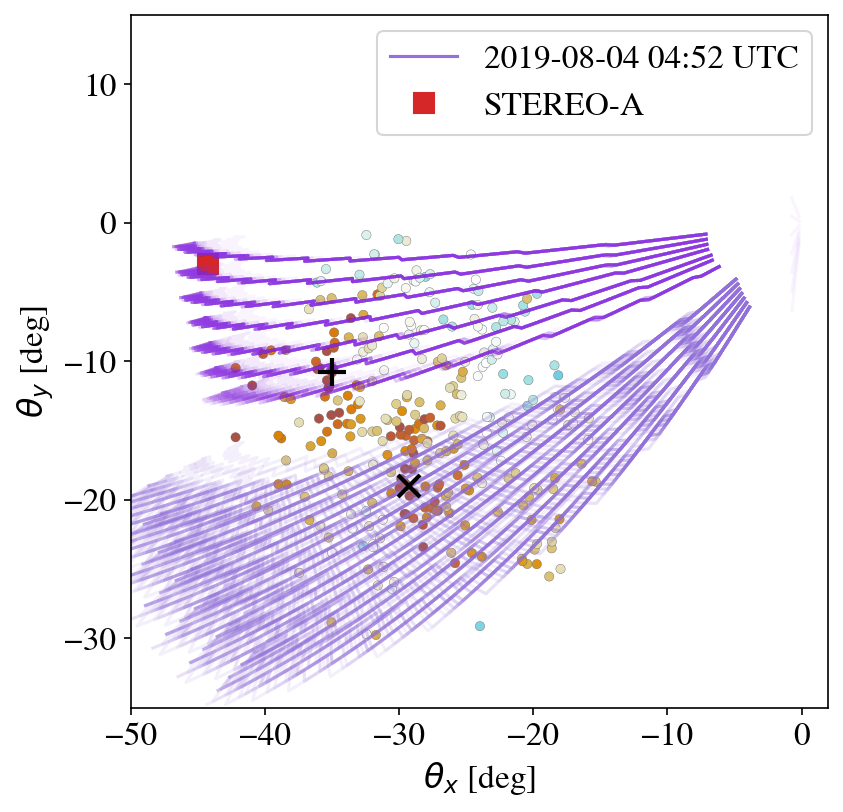}
    \caption{3D HUXt simulation of two HSS originating from an equatorial and low-latitude coronal hole, extrapolated into the MWA FOV and then projected into the HPC plane, with the location of the HSS reported for 50 ensemble runs. The HSS location is represented by the highest velocity stream for each latitude (purple lines) at the specified timestamp of 04-Aug-2019 04:52 UTC (nearest to Event B observation time). The centre point of both the upper and lower enhancements and the location of STEREO-A are included for reference. We compare the resultant streams to the original IPS g-level measurements.}
    \label{fig:sir}
\end{figure}

Following a similar procedure to that of Event A, the location of the HSS was mapped through the heliosphere by tracking the region of the stream with the highest velocity for each latitude in the simulation, and then projecting the resulting stream into the Helioprojective coordinate frame. We show the results of a single timestamp, at roughly the time of the MWA observations, in Figure~\ref{fig:sir}. The lines represent the full path of the HSS at specific moment of the timestamp. As done previously, we compare the location of the simulated event with the g-level measurements of Event B.

For both the upper and lower arcs, the centre of the enhancements lie towards the edge of the simulated equatorial and low-latitude HSSs respectively. Evident in the bottom panel of Figure~\ref{fig:huxtstream} and in Figure~\ref{fig:sir} is that over multiple ensemble runs, STEREO-A is impacted by the HSS created by the equatorial coronal hole as was observed by PLASTIC and IMPACT, which is noted as a SIR in the STEREO-A events list. 
It appears that the equatorial HSS creates a SIR that is then recorded by STEREO-A, and we see that the location of the same HSS intercepted with the IPS enhancement. Although there is very good overlap between the HSS from the low-latitude coronal hole and lower arc, the same does not apply for the upper arc and equatorial coronal hole. It is important to note that only the very centre of the HSS will be traced in Figure~\ref{fig:sir}, whereas the enhanced turbulence showing up strongly in IPS may be created at the edge of the stream, not the centre. 

Although further modelling is required to conclusively link this event with the coronal holes observed on the solar disc, it appears that as was concluded by \citeA{waszewski2023}, the lower enhancement region in Event B is being created by a SIR originating from a low-latitude coronal hole. As has been revealed by this analysis, the upper enhancement may have similar origins. The IPS observations may have also captured a level of interaction between these two stream events that may not be encapsulated by the HUXt simulations, resulting in the upper arc. Event B appears to be a density enhancement event within the heliosphere that has not been fully characterised by commonly-used identification techniques as it was moving well out of the ecliptic. Although Events A and B are relatively similar in how they present themselves in the IPS data, since Event A travels closer to the ecliptic it was listed in heliospheric catalogues, whilst Event B was not.

\section{Discussion}
\label{sect:discuss}
Although the number of reported events (32 IPS enhancement events and 79 catalogued events) is quite small, we begin to see trends arise from the comparison between IPS and other identification techniques. This simple analysis of crossmatching detected events, allows us to analyse the catalogues themselves, as well as form a better understanding of what IPS measurements are able to identify in the heliosphere.

Through the identification of both a CME in Event A and two SIRs in Event B, we have strengthened the argument that IPS observations are not just sensitive to CME shocks, but capture several sources of density shocks in the heliosphere \cite<e.g.>{Bisi2010, Tokumaru2023EW}. This sentiment is further supported by the catalogue comparison. Despite the low number of IPS event matches in the catalogues (17 out of 32), the IPS observations had a strong  overlap with the STEREO-A events list, which reports on both CME and SIR events.

We have shown that the MWA IPS observations have the additional capability of regularly identifying events that are travelling out of the ecliptic. Over 55\% of IPS enhancements were located more than 10$^\circ$ away from the solar ecliptic. These interplanetary events would not have been detected by satellites, such as STEREO-A. As it is clear that IPS measurements also detect SIRs, which are not able to be detected via white-light coronagraphs, these out-of-ecliptic events will greatly contribute to the lower success rate of crossmatching IPS events back to the catalogues as compared to identifying catalogued events in IPS observations (53\% and 68\% respectively). 
It is also evident from the identification of the two example IPS events. Both Events A and B appear relatively similar in the IPS observations, with similar position angles, but only one of the two events was detected in the catalogues. It happened to be that only the CME event was catalogued due to white-light coronagraph images, whilst the SIR was missed. It is therefore trivial to assume that other similar IPS enhancements may have also been undetected via commonly-used methods. 

The CME that was captured in Event A was only catalogued as it was captured in the coronagraph images of SOHO LASCO, with no measurements of it after it left the coronagraph's FOV. The MWA IPS observations were able to provide observations of this CME well into the heliosphere. For Event B, only one of the two SIRs were catalogued, as the only means of detecting such shocks is through in-situ measurements. It has been shown that the space weather environment was dominated by contributions from low-latitude coronal holes and SIR events in cycle 24 \cite{Luhmann2022, Tokumaru2023EW}, which created several co-rotating interaction regions (CIRs; SIRs that co-rotate with the Sun) that have been recurrent with the solar rotation through 2018 and 2019 \cite{Hudson2021}. These CIRs are not dissimilar to the SIRs captured in Event B. It is only with the inclusion of the MWA IPS observations that we were able to identify this out-of-ecliptic SIR.

\subsection{Comments on commonly-used identification techniques}
Catalogues themselves will have their own strengths and weaknesses in detecting and characterising events, with none of the catalogues having full overlap with another. The CDAW SOHO LASCO CME catalogue and the CACTus CME catalogue are both based on the same white-light coronagraph images taken by the LASCO instrument onboard SOHO, with the only differences between them being the method of CME detection (manual vs automatic) and the method of determining CME velocity, making event comparison between the two non-trivial. Comprehensive comparisons between these catalogue were reported earlier \cite{Yashiro2008}, although our analysis in this context brings a new and independent perspective that would help in planning future validation studies of catalogued heliospheric transient.
We found that 80\% of the events reported in the CACTus CME catalogue were also listed in the CDAW SOHO LASCO CME catalogue, whilst in reverse only 57\% of CDAW events were reported in CACTus. It was found that for all the IPS enhancement events that matched with an event reported in CACTus (4 events), were also matched with the same event identified in the CDAW catalogue (see Figure~\ref{fig:pie}). This is the case despite the crossover between CACTus and CDAW only making up a small percentage of all CDAW matches (36\%). It's clear from the previous results that the crossover between the CACTus and CDAW catalogues is quite substantial, where CDAW appears to encapsulate the majority of CACTus identified events.
The discrepancy between the two catalogues may be caused by the additional filtering process that was applied to the CACTus catalogue, which specifically filtered out narrow and weak CMEs. CACTus has been shown to identify more narrow CMEs than the CDAW catalogue, particularly during periods of decreased solar activity \cite{Yashiro2008}. If a similar filtering process was applied to the CDAW catalogue, or was removed from the CACTus catalogue, there may be a larger agreement between the two catalogues.

Although commmonly-used methods of observing a CME as it is ejected from the corona in white-light images will be able to capture CMEs at all position angles, the same cannot be said for SIRs. As SIRs are restricted to interplanetary space, the main way of detecting and cataloguing SIRs are through the use of in-situ measurements taken at the ecliptic, such as at a spacecraft like STEREO-A through the use of its instruments PLASTIC and IMPACT, or from measurements made at Earth. 

Ulysses \cite{Wenzel1992} was a successful solar mission, which was able to make measurements of S/CIRs not only beyond the 1\,AU bounds of Earth, but well out of the ecliptic \cite{Gosling1993b, Gosling1995}. Since the decommissioning of Ulysses, we have been left with several questions about SIRs still to be answered, one of which is being able to investigate the latitudinal structure of S/CIRs out of the ecliptic within the inner heliosphere ($<$1\,AU) \cite{richardson2018}.
Parker Solar Probe \cite<PSP;>{Fox2015} has been designed to cover a large portion of the inner heliosphere, being able to conduct close fly-bys of the Sun. Since its launch, PSP has now conducted its own first release of its S/CIR catalogue \cite{Allen2021}, which recorded two SIR events during our 2019 sampled time period during one of its fly-bys. However, PSP remains within several degrees of the ecliptic during its orbit \cite<e.g.>{Rivera2024}. 
Solar Orbiter \cite<SolO;>{Muller2020} which launched in 2020 has also begun to identify SIRs and their related properties \cite{Allen2025} within the inner heliosphere. SolO has the additional benefit of having incremental inclination changes introduced to its orbit, allowing it to reach higher solar latitudes, far out of the ecliptic \cite{GarcaMarirrodriga2021}. Another spacecraft that is on track to provide out-of-ecliptic information is the Polarimeter to UNify the Corona and Heliosphere \cite<PUNCH;>{DeForest2022} mission, which has onboard a narrow and wide-field imager with aims to understand how interplanetary events evolve through the solar wind.

We have already shown how the MWA can be used to identify out-of-ecliptic events with great detail, therefore it is clear that PUNCH and the MWA will have particularity strong synergy, as both instruments will have a similar FOV, providing the first opportunity to directly compare white-light images to radio data.
All these spacecraft will provide fundamental in-situ and remote imaging information about how S/CIRs and other interplanetary events evolve in the inner heliosphere, which in tandem with IPS observations can potentially start to answer unknowns about SIRs. This sentiment can also be expanded to include the detailed tracking of CME evolution in the inner heliosphere.

\subsection{Limitations of this study}
It is clear that IPS observations are sensitive to all types of density enhancements that exist within the heliosphere. As there is currently no clear way to differentiate between the sources of density enhancements from just the IPS measurement, it raises the potential situation of a detected IPS event being wrongly associated with a particular catalogued event. During the solar minimum period, we do not expect this situation to occur frequently, as heliospheric events become less frequent and share less overlap between neighbouring events (both in time and space), therefore the likelihood of an IPS event being matched to multiple catalogued events is quite low for this study.
There is also the potential that an IPS enhancement event that has been created by a faint, uncatalogued event has been wrongly associated with a different, catalogued event. Again, although this situation is possible, the low solar activity would decrease the chances of crossover in events. We believe the discrepancy between IPS and catalogued events is dominated by the out-of-ecliptic events.

There are also several limitations that are introduced from the nature of MWA observing, that may be introducing discrepancies into crossmatching IPS events with catalogued events. 
IPS observations taken by the MWA are designed to maximise spatial and temporal coverage. To achieve the highest level of sensitivity the 6 10-minute daily pointings must be taken clustered around local noon over a range of $\sim$\,6 hours. These pointings will usually be scheduled in order of pointing, starting with observations on the west and moving to the east. Although this allows for maximising spatial coverage, there is always the chance that a catalogued event may have been missed by the daily observations. Due to the large size of the MWA FOV, anything travelling in the plane-of-the-sky at less than 450\,km/s ($\sim1^\circ$/hr) will remain in the MWA FOV for least 24 hours. This reduces the likelihood of slower events being missed by the daily observations, but any feature moving faster than this could be passed over.
If IPS observations were used to track these features through the heliosphere, we are able to mitigate for the effects of timing by including g-level measurements made in different parts of the world, such as those collected by ISEE \cite<Institute for Space-Earth Environmental Research; e.g.>{Iwai2021, Tokumaru2023SC} and LOFAR \cite<Low-Frequency Array; e.g.>{Fallows2012, Jackson2023}. To investigate the potential for such cross-collaboration, joint studies between LOFAR and the MWA are planned for the near future.
There is also the additional effect of the sensitivity and location of the piercepoint, and the half-power region of the line-of-sight measurement. IPS measurements will be weighted to the inner heliosphere that is closest to the Earth, with the edge of the half-power region reaching only to the direct east and west limbs. It will therefore not be at its peak sensitivity for events that are travelling away from Earth, in particular weaker/less dense events. 

It should also be noted that another possible limitation of this study is assuming the radial propagation of CMEs, especially during the analysis of Event A. CMEs are known to suffer deflections in the lower corona \cite[i.e.]{Gopalswamy2009deflect, Wang2020}, therefore altering their trajectory in interplanetary space \cite{Lugaz2012}. This would introduce propagation errors when extrapolating out a CME into the MWA FOV using the assumed drag-based model. These limitations can be avoided in future studies by using 3D reconstruction techniques to estimate the actual propagation speed and direction of the CME, rather than relying on the HUXt models provided.

\subsection{Future Work}
As previously mentioned, near future plans see the cross-collaboration of several IPS-capable telescopes. IPS observations made by a global network have the added advantage of probing any position angle around the Sun when required. A current issue in heliophysics modelling is the lack of full coverage and accuracy in measurements of the solar and heliospheric environment \cite{Cheung2023}, particularly in keeping an updated model of the solar wind and solar disc rotating from the east. Several satellites, such as STEREO-A and Solar Orbiter, can only provide information about the eastern limb at certain points in their orbits, whereas IPS is able to provide around-the-clock coverage. Before the low-latitude coronal hole was visible in satellite images taken around Earth (such as the SOHO AIA imager, or STEREO-A), the IPS observations would be able to be fully processed and analysed. 

IPS observations may also be useful in further studies of the origins of the fast and slow solar wind. It was shown for Event B, that a large portion of the half-power region of the line-of-sight for the lower arc directly crosses over the low-latitude coronal hole, intersecting two sides of the coronal hole boundary, whilst for the upper arc, the line-of-sight skirts the boundary of the equatorial coronal hole. It was also shown by the HUXt model runs, that the piercepoints for both enhancements lined the edges of the simulated HSSs. A SIR will be created as a HSS pushes into the surrounding, slower solar wind. The enhancements lying on the edge of the streams in the ecliptic plane, and crossing the coronal hole boundaries in the projected plane potentially show how this density interaction region is created on either side of the HSS. Slow solar wind is typically associated with coronal hole boundaries \cite{Abbo2010, Brooks2015, Macneil2019}, although these boundaries can be difficult to define \cite{Reiss2021}. Further modelling of the coronal holes that were associated with Event B must be undertaken to investigate how this boundary could cause an interaction region in the inner heliosphere, specifically any open or closed magnetic field lines associated with it and its boundary \cite{Aslanyan2022}.

We have also shown a small example in this work of how IPS observations can be directly compared to in-situ observations made by solar spacecraft, such as STEREO-A. When these spacecraft are in the MWA's FOV, especially if they are close to the piercepoint, we are able to directly compare magnetic field and solar wind density measurements with IPS measurements. To strengthen the link between in-situ and IPS data, future work should be conducted in finding and scheduling crossover observations with the variety of spacecraft that are available. The MWA has a plethora of archived processed data which will have observations that capture spacecraft within the FOV, and for satellites such as PSP and SolO may be near the piercepoint or the half-power region. 

\section{Conclusions}
In this study, we have conducted a comprehensive comparison of MWA IPS data and several heliospheric event catalogues over a 7-month period spanning solar minimum. We have found that 57\% of all catalogued events, including CMEs and SIRs, can be identified in IPS observations. By only including the 84\% of catalogued events that had MWA data available, this increases to 68\% of events that were captured in the MWA IPS observations. 
Conducting the reverse analysis, over 53\% of enhancement events identified in the MWA IPS observations can be, with some degree of certainty, related back to a catalogued event. It is clear that the IPS signature is sensitive to all types of density fluctuations that may be present in the heliosphere, not exclusive to CMEs, but instead a mixture of CMEs and SIRs.

A significant percentage of heliospheric events identified within the MWA IPS observations were situated more than 10$^\circ$ out of the ecliptic. Two such events, Events A and B, were selected for further analysis as they represented catalogued events that were only marginally detected with commonly-used identification methods. The first enhancement event, Event A, was shown to be a small CME, which aligned with a catalogued event recorded in the CDAW SOHO LASCO CME catalogue. Event B, the second of the two events, has strong associations with two SIRs potentially originating from equatorial and low-latitude coronal holes. Although one of these SIRs was reported in the STEREO-A Events list, the other was not. We predict that such a situation is common, where several out-of-ecliptic events captured by MWA IPS observations may have been missed via commonly-used techniques.

Through this work, we have demonstrated the capability of the MWA to identify both CMEs and SIRs in IPS observations, as well as having the additional benefit of being able to probe large sections of the inner heliosphere, even out of the ecliptic. Although white-light images are able to sample all solar latitudes, they are restricted to observations very near the Sun, and the current methods of identifying SIRs rely on in-situ measurements made by spacecraft or at the Earth. The inclusion of MWA IPS observations can potentially greatly increase the number of heliospheric events that can be identified and characterised, as well as begin to shed light on unanswered questions about the structure and evolution of the solar wind in the inner heliosphere.

\section*{Open Research}
\label{open}
MWA data is available from the MWA All-Sky Virtual Observatory \cite{ASVO}, and for this work was accessed via giant-squid \cite{giant}, which is an alternative MWA ASVO client. For access to the data stored in this archive, registration is required. At the time of writing, the observations used in this paper are public, and can be identified by their GPS start times which serve as unique identifiers of these observations within the MWA archive. All the IPS observations described in \citeA{ipssurvey} are also archived, under project code \textsc{D0011}.
This research used version 4.0.2 \cite{sunpy} of the SunPy open source software package \cite{sunpy_community2020} for coordinate conversions, and the use of \textsc{matplotlib} \cite{Hunter2007} for all visualisation. 
This work relied on the use of \textsc{topcat} \cite{2005ASPC..347...29T, 2006ASPC..351..666T}. 

\acknowledgments
This scientific work makes use of Inyarrimanha Ilgari Bundara, the Murchison Radio-astronomy Observatory operated by CSIRO.
We acknowledge the Wajarri Yamatji people as the Traditional Owners of the Observatory site.
Support for the operation of the MWA is provided by the Australian Government (NCRIS), under a contract to Curtin University administered by Astronomy Australia Limited.
We acknowledge the Pawsey Supercomputing Centre which is supported by the Western Australian and Australian Governments.
A.W was supported by an Australian Government Research Training Program (RTP) Stipend and RTP Fee-Offset Scholarship, as well as a CSIRO Top-Up and Student Support Scholarship.
This work made use of several pre-complied CME catalogues. The SOHO LASCO CME Catalog is generated and maintained at the CDAW Data Center by NASA and The Catholic University of America in cooperation with the Naval Research Laboratory. SOHO is a project of international cooperation between ESA and NASA. This paper also uses data from the CACTus CME catalog, generated and maintained by the SIDC at the Royal Observatory of Belgium.
This research was funded in whole or in part by the Austrian Science Fund (FWF) [10.55776/P34437]. For open access purposes, the author has applied a CC BY public copyright license to any author-accepted manuscript version arising from this submission.

\section*{Appendix}


\subsection*{A: CME Arrival Time Drag-Based Model}
Detailed below are the equations used to calculate the evolution of a particle's velocity within the solar wind, assuming completely radial flow. This model was used to calculated the time of lift-off for an event leaving the solar surface and travelling out to the IPS FOV. 

Any particle moving outwards from the Sun in the heliosphere will experience instantaneous deceleration, $a$, for a given velocity, $v$, 
\begin{equation*}
    a = -\gamma(v-w)|(v-w)|
\end{equation*}
where $w$ is the background solar wind speed and $\gamma$ is the drag factor \cite{Cargill2004,Vrnak2012}.

For this model a background solar wind of 350\,km/s was used, with a drag factor of $10^{-11}$ \cite{Vrnak2012}.

Assuming a starting distance of one solar radii, we find the distance travelled to the IPS FOV. 
\begin{equation*}
    s = s_{IPS} - s_{surface}
\end{equation*}

Therefore, assuming radial flow, the time taken for a particle to travel distance, s, is given by
\begin{equation*}
    t = \frac{\sqrt{2as + v^2} - v}{a}
\end{equation*}

We examine the effects of changing the base ambient solar wind speed used for this analysis by using the test case of the CME captured in LASCO C2 images as described in Section~4.1. We vary the ambient solar wind speed between the range of 300--700\,km/s \cite<typical solar wind speeds measured in-situ at 1\,AU, >{Bunting2024}, noting the time of arrival into the MWA FOV for each in Table~\ref{tab:testvelocity}.

For the maximum speed of 700\,km/s we find the time of arrival varies by 10 hours to the time found for 350\,km/s. Despite this time difference, as the MWA FOV is so large, something travelling at 700\,km/s would remain in the FOV for over 12 hours. Compared to the time of the actual MWA observation on the 04-Jul-2019 07:17 UTC, it is over 12 hours in difference, although it could still be within the FOV.
In the case of typical solar wind speeds for solar minimum (300--500\,km/s), the time of arrival only varies by 3 hours between them, ranging from $\sim$3--7 hours difference to the actual MWA observation, all still well within the FOV.

It has been shown that the assumption of the ambient solar wind velocity does not introduce significant uncertainty into the time-of-arrival of events from the Sun. Therefore, we opt to use the median value stated in \citeA{Vrnak2012}.

\begin{table}[]
    \centering
    \begin{tabular}{cc}
    \hline
    \makecell[c]{Ambient solar\\wind speed\\(km/s)} & \makecell[c]{Time of arrival\\UTC} \\ \hline
        300 & 04-Jul-2019 03:40 \\
        350 & 04-Jul-2019 02:47 \\
        400 & 04-Jul-2019 02:34 \\
        450 & 04-Jul-2019 02:00 \\
        500 & 04-Jul-2019 00:48 \\
        600 & 03-Jul-2019 21:07 \\
        700 & 03-Jul-2019 16:51 \\ \hline
    \end{tabular}
    \caption{The time of arrival into the MWA FOV as extract from the DBM of a CME travelling at 384\,km/s first observed in a LASCO C2 image on the 01-Jul-2019 19:12 UTC for different assumed ambient solar wind speeds. The MWA observation was taken on 04-Jul-2019 07:17 UTC.}
    \label{tab:testvelocity}
\end{table}

\bibliography{references}

\end{document}